# High-channel-count 20 GHz passively mode locked quantum dot laser directly grown on Si with 4.1 Tbit/s transmission capacity

SONGTAO LIU,[1,†,*] XINRU WU,[1,2,†] DAEHWAN JUNG,[3,†] JUSTIN C. NORMAN,[4]
MJ KENNEDY,[1,] HON K. TSANG,[2] ARTHUR C. GOSSARD,[1,3,4] JOHN E. BOWERS,[1,3,4]

[1]*Department of Electrical and Computer Engineering, University of California, Santa Barbara, California 93106, USA*

[2]*Department of Electronic Engineering, The Chinese University of Hong Kong, Shatin, New Territories, Hong Kong SAR, China*

[3]*Institute for Energy Efficiency, University of California, Santa Barbara, California 93106, USA*

[4]*Materials Department, University of California, Santa Barbara, California 93106, USA*

*\*Corresponding author: stliu@ece.ucsb.edu*

**Low cost, small footprint, highly efficient and mass producible on-chip wavelength-division-multiplexing (WDM) light sources are key components in future silicon electronic and photonic integrated circuits (EPICs) which can fulfill the rapidly increasing bandwidth and lower energy per bit requirements. We present here, for the first time, a low noise high-channel-count 20 GHz passively mode locked quantum dot laser grown on complementary metal-oxide-semiconductor compatible on-axis (001) silicon substrate. The laser demonstrates a wide mode locking regime in the O-band. A record low timing jitter value of 82.7 fs (4 – 80 MHz) and a narrow RF 3-dB linewidth of 1.8 kHz are measured. The 3 dB optical bandwidth of the comb is 6.1 nm (containing 58 lines, with 80 lines within the 10 dB bandwidth). The integrated average relative intensity noise values of the whole spectrum and a single wavelength channel are − 152 dB/Hz and − 133 dB/Hz in the frequency range from 10 MHz to 10 GHz, respectively. Utilizing 64 channels, an aggregate total transmission capacity of 4.1 terabits per second is realized by employing a 32 Gbaud Nyquist four-level pulse amplitude modulation format. The demonstrated performance makes the laser a compelling on-chip WDM source for multi-terabit/s optical interconnects in future large scale silicon EPICs.**

## 1. INTRODUCTION

Driven by the huge demand for high-performance computing and large-scale data centers, photonic interconnects employing wavelength division multiplexing (WDM) are evolving fast as it is a feasible technology to meet the high-bandwidth and low energy consumption requirements required for data centers [1–4]. Silicon, as a major photonics platform, benefiting from the mature complementary metal–oxide–semiconductor very-large-scale integration (CMOS-VLSI) fabrication technology, offers a great opportunity to realize a low-cost, small footprint, highly scalable and energy-efficient solution. Large scale silicon electronic and photonic integrated circuits (EPICs) for inter-chip data communications have been demonstrated with unprecedented performance [5,6]. Further breakthroughs in bandwidth density and total power consumption require a full integration strategy where all the photonic components should be integrated on chip, including modulators, photodetectors, (de)multiplexers, polarization rotators, and especially, on-chip light sources with a low power consumption and high channel count for efficient WDM [7]. While the most aforementioned components can be easily realized on silicon with high performance, the on chip light sources, on the other hand, are rather difficult to realize due to the inherent indirect-bandgap nature of silicon [8]. Research has been devoted to demonstrate highly efficient and reliable WDM sources on Si, either by flip-chip bonding [9] or wafer bonding [10–12] technique. While the device performance of these lasers are comparable to or even superior to equivalent lasers on native substrates, they are arguably not the optimum choices in terms of cost, yield, scalability and reliability [13]. Recent breakthroughs on direct growth of InAs quantum dot (QD) materials on CMOS compatible on-axis (001) silicon substrates provide an appealing direction from both a cost and performance perspective, as it combines both advantage of QD active material and silicon mass production ability [13]. Record high performance Fabry-Perot (FP) QD lasers with a lowest lasing threshold and longest lifetime of more than a million hours [14] have been demonstrated recently due to the significant reduction in threading dislocation density (TDD) in the GaAs buffer layers [15]. It is, therefore, attractive to realize QD on-chip WDM sources for future large scale silicon EPICs to fulfill the bandwidth and power consumption requirements.

An alternative approach uses silicon based distributed feedback (DFB) laser arrays to achieve multi-wavelength channels by tuning the grating periods [16–18]. While looking back at the QD material properties, it is exciting to note that the inherent inhomogeneously broadened gain spectrum, ultrafast carrier dynamics, large gain and saturable absorber (SA) saturation energy ratio and low amplified spontaneous emission (ASE) noise level make QDs a perfect active medium for mode locked lasers (MLLs), which can also be used as light sources for WDM applications [19,20]. Compared to the DFB laser array with several tens of single-channel lasers, where multiple temperature controllers, wavelength trackers and complicated packaging architecture are needed, a single MLL can generate a wide coherent spectrum with a fixed channel spacing corresponding to the cavity length, which would greatly simplify the system topology and lower the total cost as well as the energy consumption. MLLs can also be used in a wide variety of compelling applications, including high-speed photonic analog-to-digital conversion, intrachip/interchip clock distribution and recovery, millimeter wave signal generation for radio-over-fiber applications [21–23].

In this work, we demonstrate the first 20 GHz passively mode locked quantum dot laser (QD-MLL) that is directly grown on on-axis (001) CMOS compatible silicon substrate. The InAs QD-MLL employs a chirped QD design with the photoluminescence (PL) full-width at half-maximum (FWHM) broadened to 53 meV while maintaining high optical intensity. The laser operates in the O-band with a wide mode locking range. Narrow RF linewidth of 1.8 kHz with a record low timing jitter noise of 82.7 fs integrated from 4 MHz - 80 MHz is measured. The laser also shows a wide optical bandwidth with total 80 channels within the 10 dB spectral bandwidth. The integrated average relative intensity noise values of the whole spectrum and a single channel are − 152 dB/Hz and − 133 dB/Hz in the frequency range from 10 MHz to 10 GHz, respectively. Among the total lines, 64 channels were selected to realize a 32 Gbaud Nyquist four-level pulse amplitude modulation (PAM-4) transmission with an aggregate total capacity of 4.1 Tbit/s. The measured bit error ratios (BERs) of back-to-back (B2B) and 5-km standard single mode fiber (SSMF) transmission are below the forward error correction (FEC) threshold (with 61 channels below hard-decision FEC threshold and total 64 lines below soft-decision FEC threshold). The demonstrated performance suggests the Si-based QD-MLL is a strong WDM light source candidate that can be integrated in future large-scale silicon EPICs to boost the system capacity and efficiency.

## 2. DEVICE DESIGN AND FABRICATION

The InAs QD laser growth was directly completed on an on-axis (001) silicon substrate by solid-source molecular beam epitaxy (MBE). Fig. 1(a) shows the detailed QD epitaxial structure. A high quality GaAs buffer layer was first grown on Si with a low TDD value of $7 \times 10^6$ cm$^{-2}$ by optimizing the InGaAs/GaAs strained superlattice dislocation filter layers and thermal cyclic annealing process [14]. It was then followed by a 300 nm heavily n-type doped GaAs contact layer and a 1400 nm n-type AlGaAs cladding layer. Five-layers of InAs/In$_{.15}$Ga$_{.85}$As dots-in-a-well (DWELL) active region with 37.5 nm GaAs spacers were sandwiched by upper and lower unintentionally doped 50 nm GaAs waveguides. A 10 nm $5 \times 10^{17}$ cm$^{-3}$ p-modulation doped GaAs layer was also introduced to the spacers to help improve the gain and temperature performance of the laser [24]. After the deposition of a 1400 nm p-type AlGaAs cladding and a 300 nm heavily doped p-type InGaAs contact layer, the whole growth procedure was completed. Previously, the QD layers have the same QD size distribution (the FWHM of PL is ∼ 30 meV) [20]. In this study, a chirped QD design was employed to broaden the FWHM of the gain spectrum [25] while maintaining high QD optical gain property. By varying the total thickness of the InGaAs layers in the DWELL structure from 3 nm to 7 nm, a different QD gain peak position of each layer was obtained. Fig. 1 (b) shows the normalized room temperature PL spectra of the calibration samples with single dot layers representing each layer thickness used in the laser. A total of ∼70 nm peak wavelength tuning can be obtained using this method. Fig. 1(c) shows the PL spectrum from the full QD laser structure material that was utilized in this work. The PL peak centers around 1240 nm with a broadened FWHM of 69 nm, corresponding to 53 meV, suggesting that the ensemble of five chirped QD layers effectively broadens the overall gain spectrum.

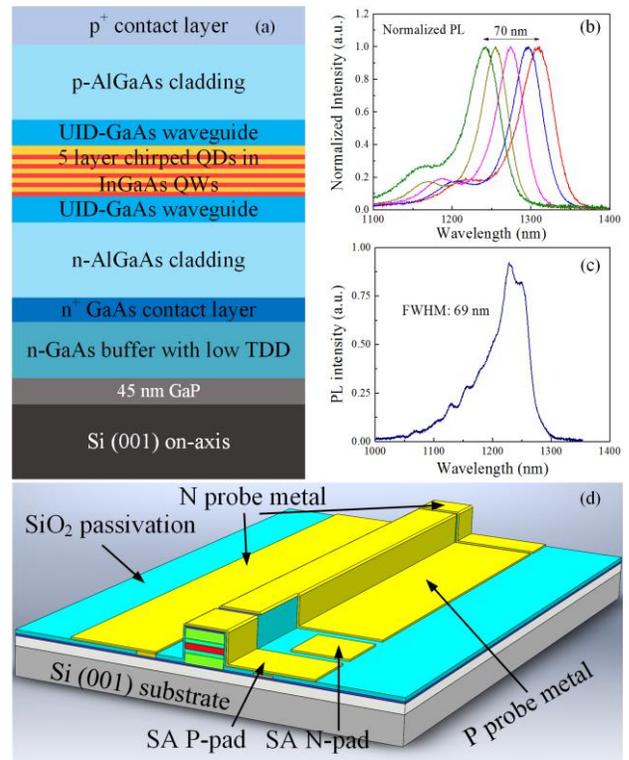

**Fig. 1.** (a) Schematic diagram of the epitaxial structure of the 20 GHz QD-MLLD (b) room temperature PL emission spectra of a single InAs DWELL layer with different InGaAs thicknesses (c) PL intensity of the full QD laser material grown and fabricated in this manuscript (d) schematic diagram of the 20 GHz quantum dot mode locked laser on silicon (not to scale).

The heteroepitaxial wafer was processed into the designed 3 μm wide deeply etched waveguide structure by standard semiconductor dry etching and metal/dielectric deposition techniques. In order to realize 20 GHz pulse generation, the total length of the laser was set to be 2048 μm (calculated by the previously obtained group effective index of 3.66). The SA section length was designed to be 14% of the total cavity length in this study. It was isolated from the gain section 10 μm away by a second dry etch step, where the heavily doped p-contact GaAs layer was etched away with ∼ 600 nm deep into the p-type cladding layer. The measured isolation resistance is around 15 kΩ. Coplanar ground-signal (GS) pads were also designed to facilitate extracting the generated RF signal directly by high speed RF probe. At the time of test, no impedance matching circuit was employed. The processed wafer was then cleaved into the designed cavity length after thinning the backside of the Si substrate down to ∼ 180 μm. The QD-MLL facets were both left as-cleaved.

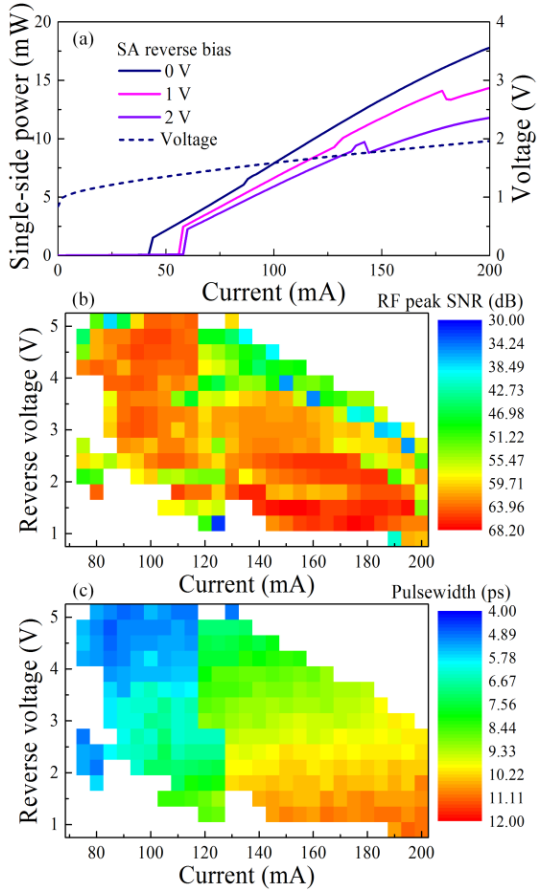

**Fig. 2.** Si-based 20 GHz QD-MLL (a) continuous wave light-current-voltage curve under different SA reverse bias (b) fundamental RF peak signal to noise floor ratio mapping and (c) pulsewidth mapping as a function of gain section current and SA section reverse bias under passive mode locking operation.

## 3. CHARACTERIZATION AND DISCUSSION

The fabricated 20 GHz QD-MLL chip was mounted on a copper heat sink and tested with a fixed stage temperature of 18°C. Continuous wave (CW) performance was first characterized as shown in Fig. 2(a). Threshold increase from 42 mA to 58 mA as the SA reverse bias voltage increase was observed for this laser due to the enhanced absorption loss within the SA section, which also caused the slope efficiency decrease. The sudden power rise at the threshold indicates the nonlinear saturation effect of the SA section as the increased intracavity spontaneous emission power would finally bleach the SA, leading to a sudden overall cavity loss drop [20]. Further optimization of the chirped QD material would help to improve the threshold current and slope efficiency. The series resistance of this sample is around 3.2 Ω. To study the passive mode locking (PML) behavior of the laser, the output spectrum was measured using an optical spectrum analyzer (Yokogawa, AQ6370C), RF performance was measured with an electrical spectrum analyzer (Rohde&Schwarz, FSU) and the autocorrelation pulsewidth was recorded with an autocorrelator (Femtochrome, FM-103MN).

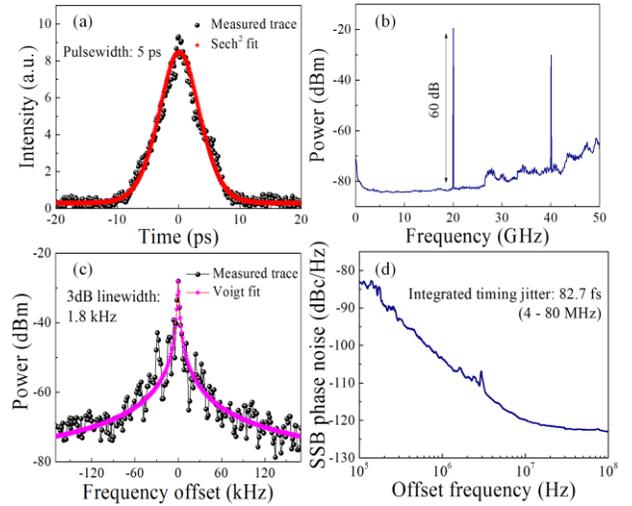

**Fig. 3.** Si-based passively mode locked 20 GHz QD-MLL (a) autocorrelation trace with hyperbolic secant squared pulse fitting (b) RF spectrum in a 50 GHz span view (RBW: 1 MHz) (c) narrow span RF peak with Voigt fit (RF peak occurred at 20.02 GHz, RBW: 1 kHz) and (d) corresponding single-sideband phase noise plot under the narrowest pulsewidth condition ($I_{gain}$ = 110 mA, $V_{SA}$ = - 5 V).

### A. PASSIVE MODE LOCKING PERFORMANCE

The PML regime of the QD laser as a function of gain section forward biased current and SA section reverse biased voltage was first delimited by defining a good mode locking state with its fundamental frequency tone signal to noise floor (SNR) ratio larger than 30 dB and corresponding pulsewidth narrower than 12 ps. A wide mode locking area was demonstrated under this criterion with forward current ranging from 75 mA to 200 mA and reverse voltage ranging from 1 V to 5 V as shown in Fig. 2(b) and 2(c) for both the SNR and pulsewidth mapping diagram, respectively. For most of the recorded mode locking states, the fundamental RF peak SNR ratio is larger than 50 dB, which indicates the good mode locking quality across the whole range. The abrupt transition around $I_{gain}$ = 120 - 125 mA is caused by the lasing mode hopping behavior, which roughly corresponds to the kink point in the corresponding light-current curves. Narrower pulses were obtained at lower current and higher reverse voltage side with high SNR values. Fig. 3(a) shows the narrowest pulse of 5 ps assuming a hyperbolic secant squared pulse profile. The pulse can be further shortened by increasing the SA section length with more effective pulse shaping dynamics within QD material [20]. RF performance is presented in Fig. 3(b). A sharp fundamental RF tone at 20.02 GHz with a SNR of 64 dB and its higher order harmonic can be clearly seen across the 50 GHz span. A battery current source as well as a linearized regulated voltage supply were then utilized to minimize the line noise in the RF linewidth measurement. Fig.

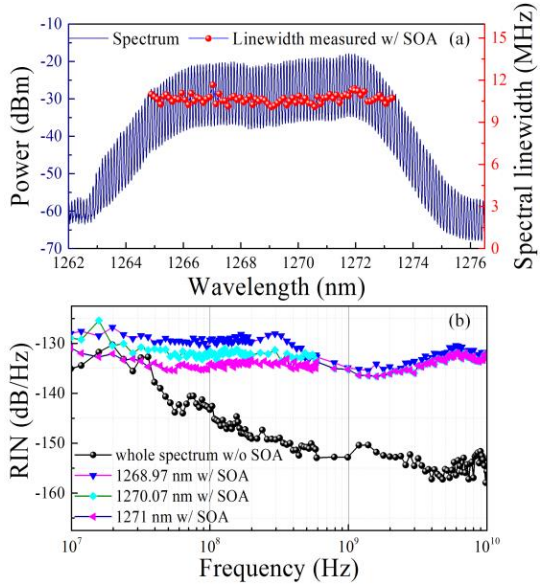

Fig. 4. Si-based passively mode locked 20 GHz QD-MLL (a) optical spectrum and corresponding optical linewidth of each mode within 10 dB (b) relative intensity noise of the whole O-band spectrum and certain filtered individual wavelength channels under $I_{gain}$ = 180 mA, $V_{SA}$ = - 1.92 V.

3(c) shows the obtained RF linewidth with a Voigt fit. The 3 dB linewidth is 1.8 kHz, which is comparable to the state-of-the-art high-speed semiconductor mode locked lasers [26,27]. Corresponding single-sideband (SSB) phase noise of this PML state is shown in Fig. 3(d). With a typical roll-off slope of 20 dBc/Hz per decade, the integrated timing jitter is 286 fs from 100 kHz to 100 MHz, and 82.7 fs from 4 MHz to 80 MHz of the ITU-T specified range, which is the lowest timing jitter ever reported to date for any passively mode-locked semiconductor laser diode. This performance is believed to benefit from the low ASE noise and low confinement factor properties of the QD material. Hermetic packaging can help to improve the stability of the QD laser further by suppressing the technical noise and electrical noise from the ambient environment that leads to better RF performance [26].

At higher pump current levels, the pulse starts to broaden to around 10 ps caused by the self-phase modulation mechanism [28]. The spectrum bandwidth also increases as more modes reach threshold due to the inhomogeneous broadening nature of the QD material. However, the SNR still maintains larger than 60 dB level on the lower voltage side, indicating strong phase correlation between these lasing modes. Fig. 4(a) shows the obtained square-like optical spectrum with largest 3 dB bandwidth of 6.1 nm (58 lines, 80 lines within 10 dB) under bias condition of $I_{gain}$ = 180 mA, $V_{SA}$ = - 1.92 V. The average optical linewidth of each mode is around 10.6 MHz measured by delayed self-heterodyne method. The corresponding SNR at this state is 66 dB with a RF 3 dB linewidth of 2.7 kHz. The relative intensity noise (RIN) as an important performance parameter was also characterized across the spectrum. As shown in Fig. 4(b), a low integrated average RIN value of -152 dB/Hz is obtained in the 10 MHz – 10 GHz range for the whole optical comb. Filtered individual channels, due to the relatively weak line power, were then amplified by an external O-band semiconductor optical amplifier (SOA) before

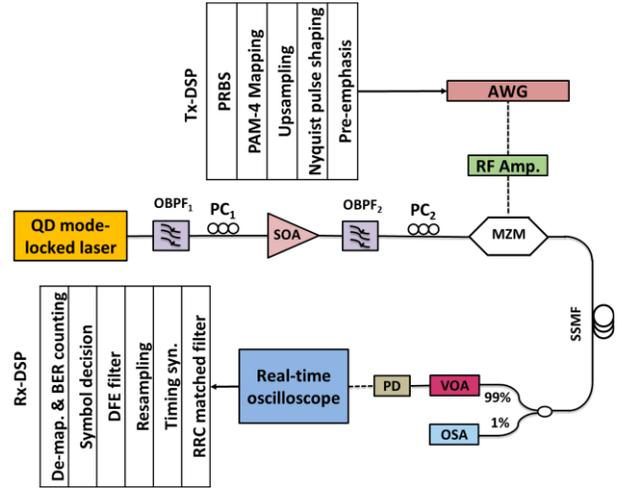

Fig. 5. PAM-4 data transmission setup, including DSP: Digital Signal Processing, PRBS: Pseudo Random Bit Sequence, BER: Bit Error Ratio, DFE: Decision-feedback Equalizer, RRC: Root-raised-cosine, PC: Polarization Controller, AWG: Arbitrary Waveform Generator, SSMF: Standard single-mode Fiber, SOA: Semiconductor Optical Amplifier, OBPF: Optical Band Pass Filter, VOA: Variable Optical Attenuator, OSA: Optical Spectrum Analyzer, PD: Photodetector.

RIN measurement. With the help of the SOA by suppressing the low frequency mode partition noise [29], the laser demonstrates an integrated average RIN value of – 133 dB/Hz of each line, which is suitable to be employed in advanced modulation format PAM-4 transmission systems to boost the bandwidth and efficiency [30].

### B. DATA TRANMISSION PERFORMANCE

Intensity modulation/direct detection (IM/DD) combined with advanced modulation format like PAM-4 was investigated due to its low complexity, high stability, and high spectrum utilization efficiency compared to coherent detection solutions, which is acknowledged as one of the promising solutions for data center interconnects [31]. To prove the suitability of this silicon-based QD-MLL for high-speed data transmission, a system-level WDM experiment employing PAM-4 modulation format with direct detection is performed. A Nyquist PAM scheme is also deployed to further improve the spectrum efficiency [32].

Fig. 5 shows the experimental setup for the Nyquist-pulse shaped PAM-4 data transmission. Each wavelength channel of the comb generated by the QD-MLL is selected by an optical bandpass filter (OBPF) (Yenista XTA-50/O) as an optical carrier. An O-band SOA (Innolume QD SOA) is followed to boost the carrier power up to ~ 12 dBm. After the SOA, another OBPF (Santec OTF-350) is used to filter out the out-of-band ASE noise. Then the amplified optical carrier signal is launched into a 30-GHz lithium niobate Mach-Zehnder modulator (MZM) (IXblue MX1300-LN-40) for data modulation. The Nyquist-pulse shaped PAM-4 symbols are generated offline by the transmitter digital signal processing (Tx-DSP) block diagram as shown in Fig. 5. The generated symbols are loaded to the arbitrary waveform generator (AWG) (Keysight M8196A, with 32-GHz bandwidth and 92 GSa/s). Then the electrical PAM-4 signals generated by the AWG is amplified to approximately 3.5 V peak-to-peak by a 38-GHz broadband RF amplifier (SHF 806E). A variable optical attenuator is added after the modulator to control the received optical power before the photodetector. Finally, the modulated optical signal is received using a 30-GHz photodetector (Finisar XPRV2022A) with a transimpedance amplifier (TIA). No optical amplifier is used after the modulator. The received signal is then sampled at 200 GS/s by a 70-GHz real-time oscilloscope (Tektronix DPO77002SX) and processed by the receiver DSP (Rx-DSP) for signal demodulation and error counting.

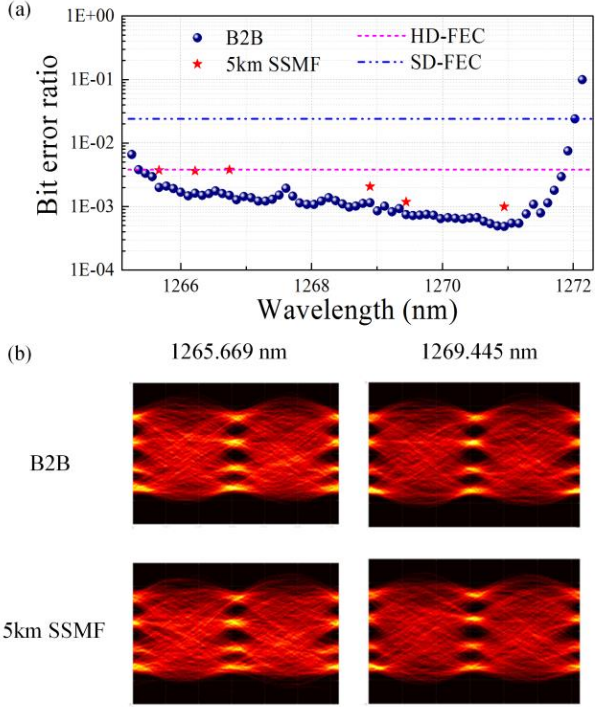

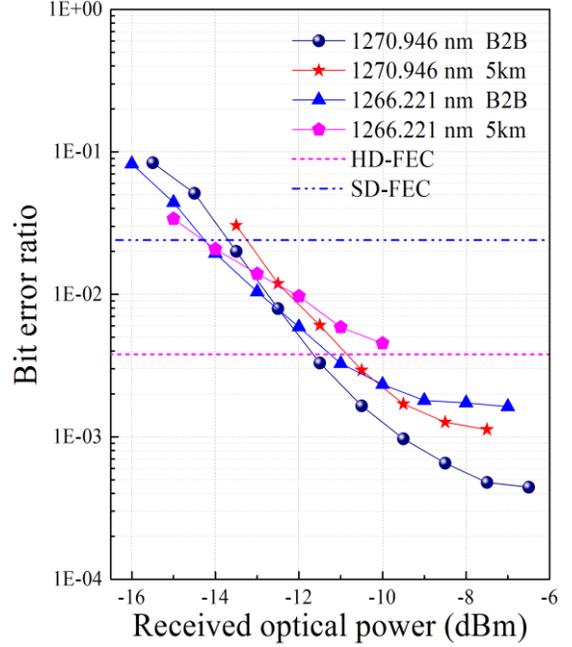

Fig. 6. (a) BER performance of the PAM-4 signal with different comb lines; (b) Corresponding eye diagrams for channels at 1265.669 nm and 1269.445 nm after back-to-back (B2B) and 5-km standard single-mode fiber (SSMF) transmission.

Fig. 7. BER versus received optical power for B2B and after 5-km SSMF transmission using the comb lines located at 1266.221 nm and 1270.946 nm.

At the Tx-DSP side, the 32 Gbaud PAM-4 signal is generated and up-sampled by a factor of 2.875 (92/32). A root-raised-cosine (RRC) filter is applied with the roll-off factor of 0.12 for Nyquist filtering. Pre-emphasis is then performed to compensate the system end-to-end frequency response using the measured inverse response function of the whole system. A 10% clipping is performed to reduce the high signal peak-to-average power ratio (PAPR) induced signal distortion after RF amplifier. At the Rx-DSP side, a matched RRC filter is executed to mitigate the effects of white noise and the received signals are resampled to two samples per symbol for the receiver-side equalization. All the Nyquist filtering is performed by the DSP. A conventional (33, 9) - tap T/2 spaced decision-feedback equalizer (DFE) is applied to restore the signal. Clock recovery is carried out to compensate any sampling phase and frequency offset that may exist between the transmitter and receiver clocks. Finally, the output PAM-4 signal is decoded for bit error ratio (BER) counting. In principle, the 20-GHz channel spacing allows at most 20 Gbaud data per channel. Since the Nyquist pulse shaping is used and the roll-off factor of the RRC filter is designed to be 0.12, the effective bandwidth of the 32 Gbaud Nyquist PAM-4 signal is reduced to 19.2 GHz and the spectrum of the Nyquist PAM-4 signal became rectangular. This allows higher data transmission rate with no crosstalk from adjacent channels.

The modulated optical signal is transmitted both for back-to-back (B2B) and over 5-km standard single-mode fiber (SSMF), respectively. Fig. 6 summarizes the transmission results. Total 65 carriers out of whole comb are leveraged for the experiment (under approximately the same bias condition as shown in Fig. 4). Fig. 6(a) shows the BER performance of each channel at the maximum received optical power (between - 5.5 dBm to - 6.5 dBm for all the channels). For the 65 tested channels, we obtain 64 channels with BERs below the soft-decision forward error correction (SD-FEC) threshold (20% overhead), leading to an aggregate data transmission rate of 4.1 Tbit/s. Considering the overhead of 20% for SD-FEC, the net spectral efficiency amounts to 2.7 bit s$^{-1}$Hz$^{-1}$. When considering the hard-decision forward error correction (HD-FEC) threshold (7% overhead), we can still achieve 61 channels with BERs below this threshold, which gives a total bit rate of 3.9 Tbit/s with a net spectral efficiency of 3.1 bit s$^{-1}$Hz$^{-1}$. 5-km SSMF transmission is also performed for some certain wavelengths with the corresponding BER performance shown in Fig. 6(a). Although at O-band, the main impairment to the signal transmission comes from the fiber attenuation, the BER performance is still below the HD-FEC threshold even for the carrier near the edge (at 1265.669 nm) after 5-km SSMF transmission. Fig. 6 (b) shows the corresponding eye diagrams (after Rx-DSP) for B2B and 5km SSMF transmission of the wavelengths located at 1265.669 nm and 1269.445 nm. The eye-opening of Nyquist PAM-4 is reduced compared to conventional PAM signal due to the high signal PAPR [33]. Fig. 7 shows the BER versus received power for the carriers located at 1266.221 nm and 1270.946 nm for B2B and after 5-km SSMF transmission. The power penalties for these two measured channels are about 1 dB. BER floor can also be observed from Fig. 7 due to the insufficient eye-opening, which can be further improved by optimizing the clipping ratio of the transmitted data.

## 4. CONCLUSIONS

We have designed and presented a high-channel-count and low-noise 20 GHz passively mode locked quantum dot laser that is directly grown on a CMOS compatible Si substrate for the first time. The QD-MLL shows excellent phase and intensity noise performance. Narrow RF linewidth with record low timing jitter value of 82.7 fs (4 – 80 MHz) as well as low RIN values are demonstrated. A wide coherent optical spectrum is also shown due to the adoption of a chirped QD active region design. By employing 64 wavelength channels as optical carriers, system level terabit transmission experiment is demonstrated. Combined with the direct growth nature, the QD-MLL manifests itself as a compelling candidate as an on-chip WDM light source. Future ongoing work would include the integration of QD semiconductor optical amplifiers to further boost the comb laser power performance and provide more flexibility in system level design. As the direct growth on silicon technology gradually grows mature, we expect to see a fully integrated largescale silicon EPIC in the near future.

**Funding**. Advanced Research Projects Agency—Energy (ARPA-E) (DE-AR0000843) ENLITENED program.

**Acknowledgment**. We thank Kurt Olsson and John English for their assistance in MBE chamber maintenance. The UCSB nano-fabrication facility was used.

† These authors contributed equally to this work.